\documentclass[12pt]{article}\usepackage[hyperfootnotes=false]{hyperref}
\usepackage{epsfig}
\usepackage{float}
\usepackage{empheq}
\usepackage{bbold}
\usepackage{comment}

\usepackage[utf8]{inputenc}
\usepackage{amsmath}
\usepackage{hyperref} 
\usepackage{siunitx}
\usepackage{bm}
\usepackage{dsfont}
\usepackage{cite}

\usepackage{caption}

\usepackage{amsmath}
\usepackage{amssymb}
\usepackage{graphicx}
\newlength{\abstractwidth}
\renewcommand{\thefootnote}{\fnsymbol{footnote}}
\renewcommand{\thanks}[1]{\footnote{#1}} 
\newcommand{\starttext}{
\setcounter{footnote}{0}
\renewcommand{\thefootnote}{\arabic{footnote}}}

\newcommand{\be}{\begin{equation}}
\newcommand{\bea}{\begin{eqnarray}}
\newcommand{\eea}{\end{eqnarray}}
\newcommand{\beq}{\begin{equation}}
\newcommand{\ee}{\end{equation}}

\begin{document}



\rightline{}
\bigskip
\bigskip\bigskip\bigskip\bigskip
\bigskip

\centerline{ \bf {  Nonlinear Ringdowns as Sources and Detectors of Quantum Gravitational Waves         }}

\bigskip
\begin{center}
\bf      Thiago Guerreiro\footnote{thguerreiro@gmail.com}  \rm

\bigskip
Department of Physics, Pontifical Catholic University of Rio de Janeiro \\

Rio de Janeiro 22451-900, Brazil 

\end{center}

\begin{abstract}

Is gravity quantum mechanical? 
If so, we argue that nonlinear effects in black hole ringdowns -- notably second harmonic generation -- generates gravitational waves in non-classical states. While quantum features of these states such as sub-Poissonian statistics or entanglement could in principle be measured at interferometric detectors, the tiny coupling of gravity to matter makes this extremely challenging. Drawing on ideas from quantum optics, we instead propose that the nonlinearities in ringdowns could be used as strongly coupled detectors of quantum gravitational radiation, potentially offering a new route to probing the quantum nature of gravity.

\end{abstract}

\starttext \baselineskip=17.63pt \setcounter{footnote}{0}



\section{Introduction}

Recent years have seen an increased interest in experimental proposals aimed at detecting phenomena that require a unified view of quantum mechanics and gravity for their explanation \cite{marletto2017gravitationally, bose2017spin, carney2019tabletop, krisnanda2020observable, qvarfort2020mesoscopic, aspelmeyer2022avoid, carney2022newton, bose2025massive, marletto2025quantum}. While a complete quantum theory of gravity remains elusive \cite{dyson2014graviton, penrose1996gravity, penrose2014gravitization}, the effective field theory quantization of general relativity yields low-energy predictions that could in principle be experimentally verified. For instance, quantum gravitational waves (GWs) can occupy states with no counterpart in classical general relativity, such as the vacuum, single-graviton states, and squeezed or thermal states \cite{guerreiro2020gravity}. These states, some of which can have a macroscopic mean number of gravitons \cite{guerreiro2022quantum}, can induce noise in interferometers \cite{kanno2019detecting,guerreiro2020gravity, guerreiro2022quantum, parikh2021a, parikh2021b}, resonant bar detectors \cite{tobar2024detecting} and gravitational decoherence \cite{blencowe2013effective, kanno2021noise, bassi2017gravitational}. Detecting such effects would provide strong evidence for the quantum nature of the gravitational field. However, as emphasized in \cite{carney2024graviton, carney2024comments}, the detection of gravitational noise or graviton clicks would not rule out classical models of gravitational radiation with the same level of rigor as established by non-classicality witnesses in quantum optics \cite{clauser1974experimental}. To demonstrate the quantum nature of gravitational radiation \textit{on par} with its electromagnetic counterpart, it is necessary to violate a non-classicality witness, for example either by detecting entanglement or sub-Poissonian statistics \cite{davidovich1996sub}. Determining whether this is possible will require understanding which non-classical GW states can be generated in nature and how they can be measured. 

In quantum optics, non-classical states are generated by nonlinear processes \cite{hillery1985conservation, albarelli2016nonlinearity}. Since gravity can be strongly nonlinear \cite{scheel2014geometrodynamics}, a natural question emerges: can the nonlinearities of Einstein’s theory also give rise to non-classicality? Recent numerical simulations have confirmed the presence of nonlinear effects in black hole (BH) ringdowns \cite{mitman2023nonlinearities, cheung2023nonlinear}; see also  \cite{zlochower2003mode, nakano2007second, london2014modeling} for earlier works. Here, we propose that nonlinearities during BH ringdowns may serve as a mechanism for generating GW states differing significantly from coherent states. BH mergers could therefore source non-classical gravitational radiation \footnote{Beyond effective quantization of general relativity and the standard model, new physics could also lead to the generation of non-classical GWs \cite{dorlis2025superradiant}}. 

The prospect of generating gravitational waves in non-classical states raises the question of how to detect their quantum features. As we will show, when a quantized GW interacts with an interferometric detector, certain quadrature correlation functions of the GW's state are imprinted on the quantum state of the interferometer's optical field. In principle, these correlation functions could be measured by monitoring the optical field and enough data could be recovered to establish the violation of a gravitational entanglement witness. In practice, however, the small coupling between gravity and matter makes this kind of direct detection extremely challenging \cite{feynman2018}. As with electromagnetic waves, witnessing the quantum nature of GWs will likely require efficient detectors, so the small gravitational coupling must be overcome.

Fortunately, gravity is a cumulative effect and significant nonlinearities are expected in strongly gravitating systems \cite{unruh1984steps}. This suggests the idea of using gravitational self-interactions as a smoking gun of quantum gravitational phenomena. Again, quantum optics offers insights. The output of optical nonlinear interactions -- such as three- and four-wave mixing -- is highly sensitive to quantum properties of the parent modes \cite{shen1967quantum, kozierowski1977quantum}, notably their particle number statistics \cite{ekert1988second, olsen2002dynamical, qu1995measurements, spasibko2017multiphoton}. 
We may expect that analogous phenomena occur in quantized gravity. If so, ringing BHs can be both a source and a detector for non-classical GWs. In this scenario, quantum features of the GW states might show in the relative strengths of excited modes of the ringing BH, and read-out by earth or space based detectors via BH spectroscopy~\cite{abbott2020prospects, maggiore2020science, evans2021horizon, AmaroSeoane2023}.

In the following, we review nonlinearities in BH ringdowns, focusing on second-harmonic generation. We then show that if GW perturbations of BH spacetimes are quantized, nonlinear ringdowns can produce non-classical GW states through second harmonics. These quantum states can lead to signatures in interferometric detectors, however, direct detection is very challenging due to the weakness of the gravitational coupling to matter. Finally, we note that nonlinear GW interactions in BH ringdowns can be strong, and discuss their potential use as strongly coupled detectors of non-classical GWs. 

Readers convinced that quantum properties of GWs cannot be detected at interferometers, or those only interested in GW nonlinear interactions, can skip section~\ref{interf}.

\section{Second harmonics in ringdowns}\label{GW_SHG}

Consider quasinormal modes (QNMs) of a perturbed Kerr BH indexed by angular harmonic and overtone numbers $ (\ell,m,n) $, with amplitude denoted by $ A_{(\ell,m,n)}$. Numerical simulations \cite{mitman2023nonlinearities} show that for a typical ringdown following the merger of two BHs, over a wide range of initial mass ratios, the first-order parent mode $ A_{\omega} \equiv A_{(2,2,0)} $ sources a second-order mode with amplitude $ A_{2\omega} \equiv A_{(4,4)}^{(2,2,0)\times(2,2,0)} \propto (A_{(2,2,0)})^{2} $ and frequency $ \omega_{(2,2,0)\times(2,2,0)} = 2\omega_{(2,2,0)} \equiv 2 \omega $. The amplitude of the second-order mode is comparable to the corresponding linear $(4,4)$ component, with the ratio
\begin{equation}
    \mathcal{R} \equiv \frac{A_{(4,4)}^{(2,2,0)\times(2,2,0)}}{A_{(2,2,0)}^{2}} ,
    \label{ratio}
\end{equation}
in the range $ 0.15 \sim 0.2 $ \cite{mitman2023nonlinearities}. This corresponds to GW second harmonic generation (SHG), a three-wave mixing process also known as frequency doubling. Similar numerical results are reported in \cite{cheung2023nonlinear}, where it is also shown that the phase of the nonlinear wave is twice that of its parent plus a constant shift, also characteristic features of frequency doubling. For earlier numerical relativity works on GW nonlinearities, we refer to \cite{zlochower2003mode, nakano2007second, london2014modeling}. From now on, we will denote the linear and nonlinear modes by their frequencies $ \omega $ and $ 2\omega$, for conciseness.



\section{Hamiltonian}

Quantum mechanically, SHG can be described by the effective Hamiltonian, 
\begin{equation}
    H_{I} 
    = \kappa   a^{2}b^{\dagger} + \kappa^{*}  a^{\dagger 2} b  , 
    \label{hamiltonian}
\end{equation}
where $ \kappa $ is the coupling constant and $ a \  (a^{\dagger}), b \ (b^{\dagger}) $ denote the annihilation (creation) operators for the parent and nonlinear modes, respectively. We now show that this Hamiltonian will also emerge in a quantum description of GW perturbations in a nonlinear ringdown. 

Black hole perturbation theory is carried out by considering perturbations of the background $ g_{\mu\nu}^{(0)} $ metric, 
\begin{equation}
    g_{\mu\nu} = g_{\mu\nu}^{(0)} + \epsilon h_{\mu\nu}^{(1)} + \epsilon^{2} h_{\mu\nu}^{(2)} + ...
\end{equation}
and imposing Einstein's eqs. to leading order in powers of $ \epsilon $. In doing this, we find the Zerilli-Regge-Wheeler eqs. in the case of Schwarzschild \cite{regge1957stability, zerilli1970effective, gleiser2000gravitational} and the Teukolsky eq. in the case of Kerr BHs \cite{teukolsky1973perturbations, campanelli1999second, zlochower2003mode}. For second-order perturbations, these have the schematic form \cite{mitman2023nonlinearities},
\begin{equation}
    \mathcal{T}\Psi^{(2)} = \mathcal{S}[\Psi^{(1)}]
    \label{teukolsky}
\end{equation}
where $ \Psi^{(1,2)} $ are scalar functions from which the first- and second-order perturbations $ h_{\mu\nu}^{(1,2)} $ can be reconstructed, $ \mathcal{T} $ is a second-order differential wave operator involving spatial and temporal derivatives, and $ \mathcal{S}[\Psi^{(1)}] $ is a \textit{source term} which has a quadratic dependence on $ \Psi^{(1)} $ and its first derivatives \cite{loutrel2021second}. Note $ \Psi^{(1,2)} $ and $ \mathcal{S} $ are gauge invariant \cite{campanelli1999second}.

The perturbation eq.~\eqref{teukolsky} admits separation of variables: the angular part is expressed in tensor harmonics for Schwarzschild and spin-weighted spheroidal harmonics for Kerr, the radial part satisfies time-independent Schrödinger eqs. with appropriate boundary conditions \cite{novikov2013physics}, while the temporal part corresponds to a discrete set of damped harmonic oscillators characterized by the QNMs’ frequencies \cite{maggiore2008gravitational}. We focus on two modes with frequencies $\omega$ (component of $\Psi^{(1)}$) and $2\omega$ (component of $\Psi^{(2)}$). These correspond to the $ \ell = m = 2 $ and $ \ell = m = 4 $ modes in the case of nonlinear ringdowns. We now assume weak damping, $\gamma \equiv \Im\{\omega\} \ll \Re\{\omega\}$, which is a good approximation for rapidly rotating BHs, with zero-damping modes being the extreme case~\cite{yang2013branching,yang2013quasinormal,yang2015turbulent}, and in the eikonal limit \cite{mashhoon1985stability, cardoso2009geodesic}.

After a suitable ``projection'', the contributions from each mode can be isolated \footnote{As discussed for example in \cite{zimmerman2014quasinormal, green2023conserved}}. Quantization then leads to annihilation and creation operators for the frequency modes of $ \Psi^{(1,2)}$ \cite{unruh1974second, candelas1981quantization, iuliano2023canonical}. We denote the annihilation operators for the $ \omega $ and $ 2\omega$ modes as $ a, b $, respectively.
The source term $\mathcal{S}[\Psi^{(1)}]$ couples the eqs. of motion for $ a $ and $ b $, and since it depends quadratically on $\Psi^{(1)}$ and its first derivatives, these coupled eqs. must be derivable from a Hamiltonian containing terms \emph{at most} cubic in the set $\lbrace a, a^{\dagger}, b, b^{\dagger} \rbrace$ \footnote{In fact, the quadratic dependence of the source term on $ \Psi^{(1)}$ implies the Hamiltonian must be linear in $ b $ and quadratic in $ a$, since mode $ b $ is ``forced'' by quadratic functions of $ \Psi^{(1)}$ and its derivatives, which yield quadratic expressions in $ a $ and $a^{\dagger}$}.


The above procedure for deriving the interaction Hamiltonian for gravitational SHG was recently applied in a simplified setting in~\cite{manikandan2025squeezed}, but the same conclusions follow from the quadratic dependence of $ \mathcal{S} $ on $ \Psi^{(1)} $ and considerations of approximate energy conservation \cite{guerreiro2023nonlinearities}.

Graviton number, or energy conservation requires that the Hamiltonian we are seeking commutes with the operator $ J = a^{\dagger}a + 2b^{\dagger}b$ . This is approximatelly valid for weakly damped modes, as well as for interaction times $ t \ll \gamma^{-1}$. Writing down a sum of all possible terms up to third order in the modes' creation and annihilation operators [as imposed by the form of \eqref{teukolsky} and source term], the conservation condition $ [H,J] = 0 $ leads uniquely to \eqref{hamiltonian} up to the coupling constant and a free field Hamiltonian. See \ref{A_Energy_conservation} for details. In the gravitational context, the coupling constant $ \kappa$ depends on spatial mode overlap integrals of the interacting GWs \cite{su2017black}. For simplicity, we assume $ \kappa $ real.

From the form of the interaction Hamiltonian, we can gain intuition on the quadratic dependence of the nonlinear signal on the amplitude of the parent mode. When mode $a$ is populated by a strong coherent state, we can effectively replace the operator $a$ by the corresponding c-number amplitude $ \alpha_{\omega} \propto A_{(2,2,0)}$. This is known in quantum optics as the undepleted pump approximation \cite{garrison2008quantum}. Under this assumption, the interaction Hamiltonian becomes $ H_{I} \sim \kappa \alpha_{\omega}^{2} \ b^{\dagger} + \mathrm{h.c.} $, which acts as a generator of displacements for mode $b$. The resulting evolution produces an approximate coherent state $\vert \alpha_{2\omega} \rangle \approx e^{-iH_{I}t} \vert 0 \rangle $ in $b$, where
\begin{equation}
    \alpha_{2\omega} \sim \kappa t \ \alpha_{\omega}^{2}
    \label{amplitudes}
\end{equation}
and $ \alpha_{2\omega} \propto A_{(4,4)}^{(2,2,0)\times(2,2,0)} = \mathcal{R} A_{(2,2,0)}^2 $. We thus see that $ \kappa \propto \mathcal{R} $. Later on, in section \ref{nonlinear_detector} we will use \eqref{amplitudes} to estimate properties of the GW states resulting from nonlinear interactions. 

This coherent state picture of the nonlinear mode is a crude approximation, valid only in the undepleted pump approximation. It breaks down near $\mathcal{R} \sim 1$, when strong coupling leads to significant energy exchange between the parent and nonlinear modes. 
\section{Equations of motion}

Define $ \alpha = ae^{i\omega t} $ and $ \beta = be^{i2\omega t} $. In the interaction picture, the dynamical equations governing frequency doubling read~\cite{kozierowski1977quantum, mandel1982squeezing, olsen2001predictions},
\begin{equation}
    \dot{\alpha} = -2i\kappa \beta \alpha^{\dagger} \label{OPA1}
\end{equation}
\begin{equation}
    \dot{\beta} = -i\kappa \alpha^{2} \label{OPA2}
\end{equation}
which are nonlinear coupled differential equations. We refer to \ref{Freq_doubling_eqs} for the inclusion of damping terms. In quantum optics, it is known that these eqs. lead to squeezing and sub-Poissonian statistics of the parent and nonlinear modes \cite{mandel1982squeezing, olsen2001predictions}, which are nonclassical properties \cite{hillery1985conservation, davidovich1996sub}. Moreover, input parent coherent states undergoing SHG can also generate entanglement between the parent and nonlinear modes in the strong coupling regime, provided dissipation is not too large \cite{olsen2004continuous, grosse2006harmonic, villar2006direct, coelho2009three, dechoum2010semiclassical}. Hence, if GWs are quantized, nonlinear ringdowns generate non-classical gravitational radiation.

\section{Signatures on interferometers \label{interf}}

Experimentally determining the quantum nature of gravity at the level of rigor of quantum optics, either for single mode or entangled states, requires the violation of a non-classicality witness, e.g. a Cauchy-Schwarz inequality \cite{clauser1974experimental} or entanglement witness \cite{duan2000inseparability}. We now investigate whether this could in principle be done by monitoring the quantum optical state of an interferometric detector interacting with a non-classical GW. The short answer is yes in principle, although as we will see, in practice direct measurements are extremely challenging. This is consistent with \cite{carney2024graviton, carney2024comments}.

Following \cite{guerreiro2020gravity, guerreiro2022quantum}, we consider an optical cavity as a model interferometric detector. Starting from the linearised Einstein's equations in the Minkowski background \cite{pang2018quantum}, one can show that the interaction between the cavity field and GW modes consists of a dispersive optomechanical coupling in which the GWs formally assume the role of mechanical oscillators \cite{guerreiro2020gravity, guerreiro2022quantum}. We refer to this as the \textit{optogravitational} interaction. As usual, we will work with a discrete set of GW modes in a cubic box of volume $ V $ and take the continuum limit $ V \rightarrow \infty$ when necessary. In order to encompass the case of bipartite entangled states, we consider two GW modes, but note this can be extended to the case of single-mode non-classical states or multi-partite entangled states in a straightforward manner.

\begin{figure}[t]
    \centering
    \includegraphics[width=0.5\textwidth]{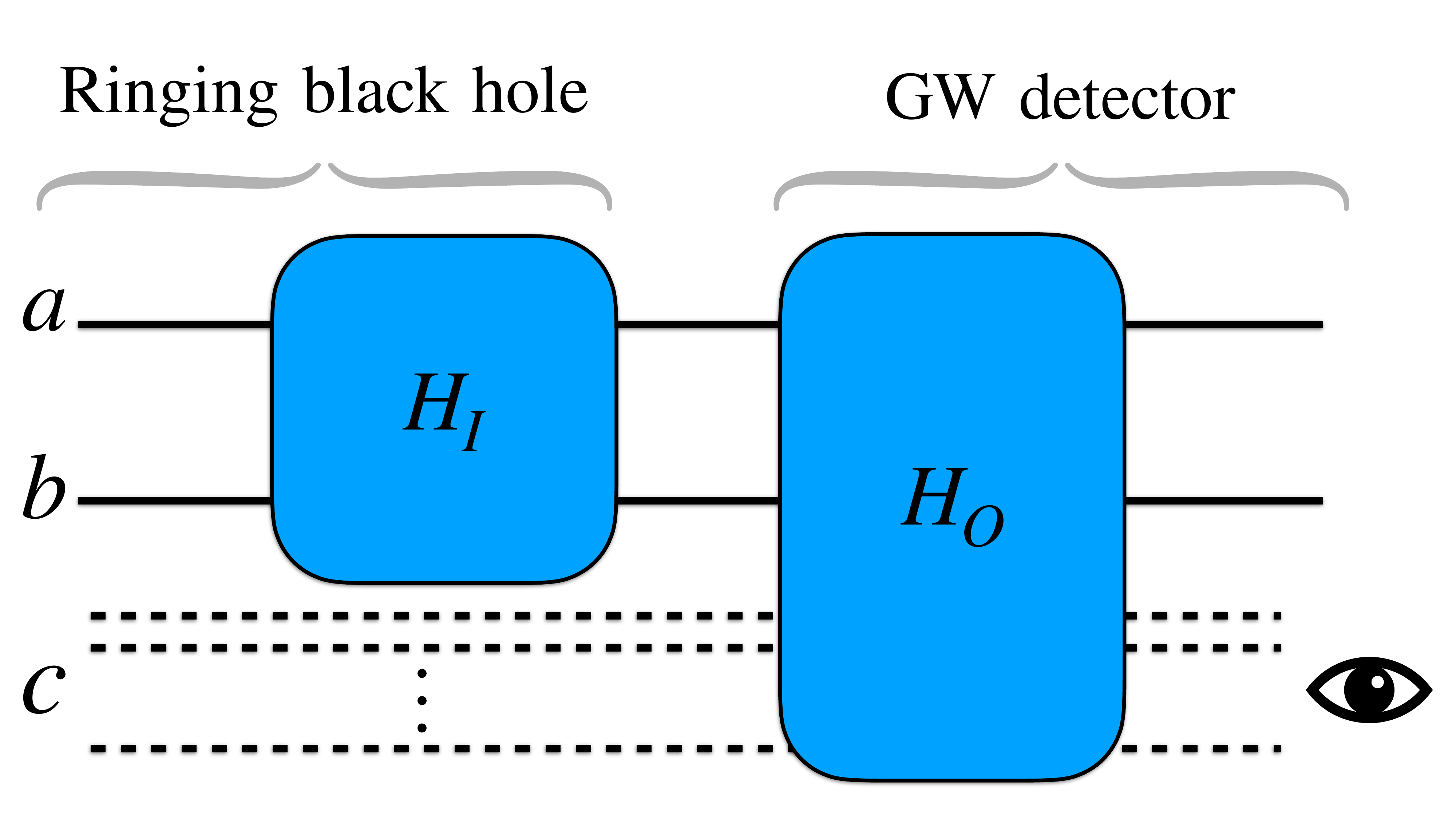}
    \caption{Concept for a gravitational wave entanglement witness. A nonlinear ringing black hole produces entangled GW states in modes $ a $ and $ b $ via the interaction Hamiltonian $ H_{I} $. The GWs subsequently interact with an ensemble of optical modes $ c $ via the optogravitational Hamiltonian $ H_{O} $. Information on quantum features of the waves, notably field quadrature correlation functions, get imprinted on the optical ensemble's state.
    }
    \label{fig1}
\end{figure}

For two GW modes $ a $ and $ b $, both in the $+$ polarization and incident perpendicularly to the cavity axis \cite{guerreiro2020gravity},  the optogravitational Hamiltonian reads,
\begin{equation}
    H_{O} = \omega_{c} c^{\dagger}c + \omega_{a}a^{\dagger}a + \omega_{b}b^{\dagger}b - c^{\dagger}c \left(  g_{a}Q_{a} + g_{b} Q_{b}   \right) \ ,
    \label{two_mode_hamilton}
\end{equation}
\noindent where $ c (c^{\dagger}) $ represents the annihilation (creation) operator for the cavity; $ \omega_{c}, \omega_{a,b} $ are the cavity and GWs' frequencies, respectively and $ Q_{a} = a + a^{\dagger}, Q_{b} = b + b^{\dagger} $ are the amplitude quadratures for the GW modes. We also define the GW's phase quadratures according to $ P_{a} = i( a^{\dagger} - a ) $ and $ P_{b} = i(b^{\dagger} - b) $. The couplings $ g_{i} $ are given by
\begin{equation}
    g_{i} \equiv \omega_{i}q_{i} = \frac{\omega_{c}}{4} \sqrt{\frac{8\pi G}{\omega_{i}V}} \ ,
    \label{coupling1}
\end{equation}
where $ i = a,b $, $ G $ is Newton's constant and we have introduced the \textit{dimensionless} coupling $ q_{i} $. Note that \cite{guerreiro2020gravity},
\begin{equation}
     q_{i} \sim (\omega_{c} / E_{\rm pl}) \ll 1 \ ,
    \label{coupling_const}
\end{equation}

\noindent which hints that GW detectors are weakly coupled optomechanical systems \cite{guerreiro2020gravity}, i.e. GW detectors are weakly gravitating systems \cite{unruh1984steps}.

The Hamiltonian \eqref{two_mode_hamilton} can be exponentiated to give the time-evolution operator, which in the interaction picture reads \cite{brandao2020entanglement},
\begin{equation}
    U(t) = e^{-iB(t)(c^{\dagger}c)^{2}} e^{q_{a}c^{\dagger}c(\eta_{a}a^{\dagger} - \eta_{a}^{*}a)} e^{q_{b}c^{\dagger}c(\eta_{b}b^{\dagger} - \eta_{b}^{*}b)}
    \label{unitary}
\end{equation}

\noindent where,
\begin{equation}
    \eta_{k} = (1 - e^{-i\omega_{k} t}) 
\end{equation}

\begin{equation}
    B(t) =  \sum_{k = a,b} q_{k}^{2} \left(   \omega_{k} t - \sin \omega_{k} t  \right) \ . \label{Bs}
\end{equation}
We refer to \cite{guerreiro2020gravity} and \ref{Optogravitational_Hamiltonian} for details.

Now, consider the GWs and optical field are initially in a separable state given by
\begin{equation}
    \rho(0) = \left(  \sum_{n,m}a_{n}a^{*}_{m} \vert n \rangle \langle m \vert \right) \otimes \sigma_{ab} \ ,
\end{equation}
where $ a_{n} $ are known complex coefficients defining the initial optical density matrix and $ \sigma_{ab} $ is the GWs' initial state. 
Using the time-evolution operator \eqref{unitary}, we can compute the reduced optical density matrix $ \rho_{c}(t) = \mathrm{Tr}_{ab} \left(U\rho_{abc}U^{\dagger} \right)$. In the photon-number basis we find,
\begin{equation}
    [\rho_{c}(t)]_{nm} \equiv 
 \langle n \vert \rho_{c}(t) \vert m \rangle = a_{n}a^{*}_{m} e^{-i \delta B(t)} e^{-i\omega_{c}\Delta t} \ \mathcal{C}_{\Delta}(t) \ ,
\end{equation}

\noindent where $ n,m \in \mathbb{N} $, $ \delta = (n^{2} - m^{2}) \ , \Delta = (n-m) $ and $ \mathcal{C}_{\Delta}(t)$ is defined as, 
\begin{equation}
    \mathcal{C}_{\Delta}(t) = \mathrm{Tr} \left( \sigma_{ab}   e^{\Delta q_{a}(\eta_{a}^{*}a^{\dagger} - \eta_{a}a))} e^{\Delta q_{b}(\eta_{b}^{*}b^{\dagger} - \eta_{b}b))}  \right) \ .
\end{equation}
Note this quantity is related to the Feynman-Vernon influence and decoherence functionals \cite{feynman2000theory, hu1992quantum, parikh2021a}.
By measuring the reduced density matrix describing the optical ensemble we can in principle obtain information on the quantum correlations between the GW modes $ a $ and $ b $. 

To quadratic order in the coupling $ q_{a,b} $,
\begin{eqnarray}
\mathcal{C}_{\Delta}(t) \approx 1 
 - i q_{a} \Delta \left(   \sin \omega_{a}t \langle Q_{a} \rangle 
   + (1 - \cos \omega_{a}t) \langle P_{a} \rangle  \right) \nonumber \\ 
 \qquad - \frac{q_{a}^{2}\Delta^{2}}{2} \Bigl(  
   \sin^{2}\omega_{a}t  \langle Q_{a}^{2} \rangle 
   + (1 - \cos \omega_{a}t)^{2} \langle P_{a}^{2} \rangle \nonumber \\
 \qquad\qquad\qquad + \sin \omega_{a}t (1 - \cos \omega_{a}t) 
   \bigl< Q_{a}P_{a} + P_{a}Q_{a} \bigr> \Bigr) \nonumber \\
 \qquad + (a \rightarrow b) \nonumber \\
 \qquad - q_{a}q_{b}\Delta^{2} \Bigl( 
   \sin \omega_{a}t \sin \omega_{b}t \langle Q_{a} Q_{b} \rangle 
   +  \sin \omega_{a}t (1 - \cos \omega_{b}t) \langle Q_{a} P_{b} \rangle \nonumber \\
 \qquad\qquad\qquad +  \sin \omega_{b}t (1 - \cos \omega_{a}t) \langle P_{a} Q_{b} \rangle 
   + (1 - \cos \omega_{a}t)(1 - \cos \omega_{b}t) \langle P_{a} P_{b} \rangle  \Bigr) ,
\label{correlator_linearised}
\end{eqnarray}
where expectation values are taken with respect to $ \sigma_{ab} $ and $ (a \rightarrow b) $ denotes the second and third terms in the first two lines with $ a $ replaced by $ b $. $ \mathcal{C}_{\Delta}(t) $ thus contains enough information to reconstruct the complete covariance matrix for modes  $ a $ and $ b $, which for Gaussian quantum states is equivalent to knowledge of the complete density matrix $ \sigma_{ab}$. 
From the covariance matrix, an entanglement witness such as for example Duan's inseparability criterion can be verified \cite{duan2000inseparability},
\begin{equation}
    \mathcal{D} = \mathrm{Var}(Q_{a} + Q_{b}) + \mathrm{Var}(P_{a} - P_{b})  \ ,
\end{equation}
where $ \mathcal{D} < 4 $ implies entanglement. 

Witnessing entanglement for GWs emitted by a nonlinear BH ringdown would establish the quantum nature of gravity, at the level of rigor of analogous experiments in quantum optics which confirm the quantum nature of light. A quantum circuit of the experiment is shown in Fig. \ref{fig1}. Unfortunately, doing this in practice is likely to be extremely hard. To see that, we turn to the question of how the optical field quadratures react to interactions with a quantized GW. 

Consider the time-evolution of the optical annihilation operator in the rotating frame $ \gamma = c e^{i\omega_{c}t} $. Its mean value is given by
\begin{equation}
    \langle \gamma(t) \rangle =  \mathcal{C}(t) \langle \gamma(0) \rangle + (\mathrm{terms \ independent \ of }\ \sigma_{ab}) \ ,
    \label{optical_annihilation}
\end{equation}
where $ \mathcal{C}(t) $ is $ \mathcal{C}_{\Delta}(t) $ evaluated at $\Delta = 1 $, i.e. $ \mathcal{C}(t) \equiv \mathcal{C}_{1}(t)$.
Defining the optical quadratures in the rotating frame as $ X \equiv \gamma + \gamma^{\dagger}$, $ Y \equiv i(\gamma^{\dagger} - \gamma) $, we find,
\begin{eqnarray}
    \langle X(t)\rangle  &\approx& \mathrm{Re}(\mathcal{C}(t)) \ \langle X(0)\rangle - \mathrm{Im}(\mathcal{C}(t)) \ \langle Y(0) \rangle\\
    \langle Y(t)\rangle  &\approx& \mathrm{Im}(\mathcal{C}(t)) \ \langle X(0) \rangle + \mathrm{Re}(\mathcal{C}(t)) \ \langle Y(0) \rangle
\end{eqnarray}
where we have ignored terms independent of $ \sigma_{ab}$ in \eqref{optical_annihilation}. For small couplings $ q_{a,b} $, this corresponds to a rotation in the optical phase space. To see that, consider for simplicity an initial GW state of the form $ \sigma_{ab} = \rho_{a} \otimes \vert 0 \rangle \langle 0 \vert_{b} $ and an initial optical coherent state with $ \langle Y_{c}\rangle = 0 $. To linear order in $ q_{a} $,
\begin{eqnarray}
    \langle X(t)\rangle  &\approx&  \langle X(0)\rangle \\
    \langle Y(t)\rangle  &\approx&  -q_{a} \left(   \sin \omega_{a}t \langle Q_{a} \rangle - \cos \omega_{a}t \langle P_{a} \rangle  \right) \langle X(0) \rangle   \label{optical_phase}
\end{eqnarray}
where we have neglected constant terms and vacuum fluctuations from mode $ b $; see \ref{Optogravitational_Hamiltonian} for a discussion. The mean phase space position of the optical field oscillates in proportion to the GW field quadratures at the frequency of the wave. The strength of the oscillations is directly proportional to the coupling constant $ q_{a}$ and the cavity quadrature $ \langle X(0) \rangle  $.

This is precisely what is observed in interferometric detectors \cite{guerreiro2020gravity}. We conclude that, to linear order in the coupling $ q_{a} $, interferometric detectors only have access to the mean field quadratures of the GW. Looking at \eqref{correlator_linearised}, we see that second moments of the GW appear at $ \mathcal{O}(q_{a}^{2})$ in $ \mathcal{C}(t)$, and hence will only affect the optical quadratures at quadratic orders in the coupling. 

Given the smallness of the coupling constant \eqref{coupling_const}, this shows that detecting second moments of the GWs would require monitoring the optical cavity with precision vastly greater than achieved in today's detectors and very likely out of reach from conceivable experiments. Measuring classical gravitational waves, which is a linear effect in $ q_{a} $, is at the limit of current technology; imagine detecting effects which are \emph{quadratic} in $ q_{a} $. This is a reflection of the weak coupling between matter to gravity, which makes GW detectors weakly gravitating systems. To detected quantum features of the gravitational field, we need strongly gravitating quantum systems. 



\section{Ringdown nonlinearities as detectors}\label{nonlinear_detector}

In quantum optics, it has long been recognized that the output of a nonlinear process strongly depends on the particle number statistics of the input parent modes \cite{kozierowski1977quantum, ekert1988second, olsen2002dynamical}. For the case of SHG, this effect can be understood through simple physical reasoning. Particle number statistics determines whether the parent fields are bunched or anti-bunched \cite{davidovich1996sub}. Energy conservation then requires the presence of two or more particles in mode $ \omega $ for the creation of a particle in mode $2\omega$. Hence, bunched states are converted more efficiently than anti-bunched ones in SHG \cite{spasibko2017multiphoton}. By analogy to quantum optics, we can therefore anticipate that the properties of nonlinearly generated GWs will depend on the graviton number statistics and other higher-order moments of the parent modes.
Given the strongly nonlinear character of gravity, these effects can be significant in ringdowns.

To quantify how strong nonlinear GW couplings in ringdowns can be, we may estimate the ``graviton conversion efficiency'' between modes, 
The mean number of gravitons in a GW of frequency $ \omega $ and amplitude $ A_\omega$ is \cite{guerreiro2020gravity},
\begin{equation}
    \langle N_\omega \rangle \approx \frac{V}{32\pi G} \omega A_{\omega}^{2}
\end{equation}
We may then define the graviton-graviton conversion efficiency during a nonlinear ringdown as 
\begin{equation}
    \eta_{\mathrm{ringdown}} = \frac{\langle N_{2\omega} \rangle}{\langle N_{\omega} \rangle} \approx 2 \mathcal{R} A_{2\omega}
\end{equation}
where we recall the gravitational wave amplitude $ A_{2\omega} \equiv A_{(4,4)}^{(2,2,0)\times (2,2,0)}$ discussed in section \ref{GW_SHG}. At the source, GW amplitudes can be of order unity \cite{yoo2023numerical}. Taking $ A^{(2,2,0)\times (2,2,0)}_{(4,4)} \sim 0.1$ and $ \mathcal{R} \sim 0.15 $  \cite{mitman2023nonlinearities} we have the conversion efficiency,
\begin{equation}
    \eta_{\mathrm{ringdown}} \sim 10^{-2} \ .
    \label{graviton_efficiency}
\end{equation}

It is interesting to contrast $\eta_{\mathrm{ringdown}}$ with the corresponding efficiencies for matter-based detectors. For instance, the conversion efficiency of gravitons into photons via the Gertsenshtein effect is approximatelly \cite{carney2024graviton, cast2017new},
\begin{equation}
\eta_{\mathrm{gertsenshtein}} \sim 10^{-33}
\end{equation}
while for bar detectors \cite{tobar2024detecting},
\begin{equation}
    \eta_{\mathrm{bar}} \sim 10^{-36}
\end{equation}


\noindent This comparison shows that ringing BHs are essentially strongly coupled nonlinear detectors, with a high conversion efficiency between different modes. Nonlinear ringdowns may therefore be used as efficient detectors for quantum properties of GWs. 

The graviton conversion efficiency \eqref{graviton_efficiency} can also be used to estimate the departure from classicality during a nonlinear ringdown. To do that, let us consider Mandel's $\mathcal{Q}$ parameter, defined as
\begin{equation}
    \mathcal{Q} \equiv \frac{\langle\Delta N^{2}\rangle - \langle N\rangle}{\langle N\rangle } \ .
\end{equation}
Classical states have $ \mathcal{Q} \geq  0 $, while $ \mathcal{Q} < 0 $ implies sub-Poissonian statistics, which cannot be described classically \cite{mandel1995optical}; $ \mathcal{Q}$ can therefore be used to distinguish classical from quantum radiation. During SHG, an initial coherent state $ \vert \alpha_{\omega} \rangle $ in the fundamental mode $ \omega$ evolves over short times to a non-classical state with \cite{mandel1982squeezing}
\begin{equation}
    \mathcal{Q} \sim - 2 \kappa^{2} t^{2} \ \vert \alpha_{\omega} \vert^{2} + \mathcal{O}(\kappa^{4}t^{4}) 
\end{equation}
where this expression is valid for $ \kappa t \vert\alpha_{\omega}\vert \ll 1$. With the help of eq. \eqref{amplitudes}, $ \kappa t \sim \vert \alpha_{2\omega}\vert / \vert \alpha_{\omega}\vert^{2}$, hence
\begin{equation}
    \mathcal{Q} \sim -2 \vert \alpha_{2\omega}\vert^{2} / \vert \alpha_{\omega}\vert^{2} \sim - \eta_{\rm ringdown} \ , 
\end{equation}
i.e. at short interaction times the fundamental mode will develop a small, but non-negligible negative Mandel parameter on the order of $\mathcal{Q} \sim -0.01$. The fundamental mode becomes non-classical. Note second harmonic also generates squeezing in the fundamental mode \cite{mandel1982squeezing}.

Let us now see an example illustrating how the statistical properties of parent states can become imprinted upon nonlinear modes in the case of SHG. Consider the dynamical eqs.~\eqref{OPA1} and \eqref{OPA2}. We take mode $ a $ to be initially populated by some quantum state $ \rho_{a} $, while $ b $ is initially in the vacuum state. For short times, the mean number of particles in mode $  b $ grows as \cite{ekert1988second},
\begin{equation}
   \langle N_{b}(t) \rangle = g^{(2)}(0)  \ \langle N_{a}(0) \rangle^{2} \  \kappa^{2}t^{2} , 
   \label{growth}
\end{equation}
where
\begin{equation}
    g^{(2)}(0) = \frac{\langle a^{\dagger}a^{\dagger}aa\rangle}{\langle a^{\dagger}a\rangle^{2}}
    \label{g2}
\end{equation}
is the second-order field auto-correlation function and expectation values are taken with respect to the initial state $ \rho_{a} $. In deriving \eqref{growth} we assumed  $ \sqrt{2g^{(2)}(0) \langle N_{a}(0) \rangle} \ \kappa t \ll 1 $ and $t \ll  \gamma^{-1} $, but note $ \langle N_{b}(t) \rangle $ also exhibits significant dependence on the parent quantum state beyond the short time or small damping approximation \cite{ekert1988second}. 

The result \eqref{growth} tells us that for short times, the efficiency with which particles are converted from modes $ a $ to $ b $ is directly proportional to the field auto-correlation function \eqref{g2}, which is related to the bunching or anti-bunching character of the wave \cite{davidovich1996sub}. For coherent states $ g^{(2)}(0) = 1 $, for thermal states $ g^{(2)}(0) = 2 $ and squeezed states have $ g^{(2)}(0) = 3 + \frac{1}{\langle N_{a}(0)\rangle} $. Importantly, for large mean number of gravitons, squeezed states have $ g^{(2)}(0) \approx 3 $, independent of $\langle N_{a}(0)\rangle$. 
Therefore, we can expect second harmonic GW generation in BH ringdowns to be significantly enhanced for parent thermal and squeezed states in comparison to coherent states.


Finally, note that other quantities associated with the nonlinear mode also carry additional statistical properties of the parent mode. Once again looking at frequency doubling, the amplitude quadrature of mode $ b $ is given to linear order in $ t $,
\begin{equation}
    \langle Q_{b}(t) \rangle \approx \frac{\kappa t}{2} \Bigl< Q_{a}(0)P_{a}(0) + P_{a}(0)Q_{a}(0) \Bigr> 
    \label{quadrature}
\end{equation}
which to leading order remains valid in the presence of dissipation; see \ref{SHG_solutions} for details. 

The most general Gaussian single-mode states consist of displaced-squeezed-thermal states defined in terms of a displacement amplitude $ h = \vert h \vert e^{i\theta} $, a squeezing parameter $ \xi = re^{i\phi} $ and mean thermal occupation number $ \Bar{n}$ \cite{weedbrook2012gaussian}. For these states,
\begin{equation}
    \frac{1}{2}  \Bigl< Q_{a}P_{a} + P_{a}Q_{a}\Bigr> = \langle Q_{a} \rangle \langle P_{a} \rangle 
    + (2\bar{n}+ 1)\sinh r \sin \phi \cos \phi 
\end{equation}
The first term on the r.h.s. is due to the coherent displacement of the GW and is the only term present for coherent states. In contrast, the second term depends on non-coherent properties of the state, the mean thermal occupation number $ \bar{n} $ and squeezing parameter $ \xi$. Note the non-coherent term is accompanied by an exponential enhancement characteristic of squeezing~\cite{guerreiro2020gravity, parikh2021b}. 
 
As we have seen, interferometric detectors are sensitive to the mean GW quadratures [see \eqref{optical_phase}]. Eq. \eqref{quadrature} is interesting, as it shows that SHG encodes information on the covariance of $ a $ in the mean quadrature of $ b $, which can be read-out by interferometric detectors. Detection of $ \langle Q_{b}(t) \rangle $ by standard GW observations could therefore signal deviations of mode $ a $ from a coherent state. For a discussion on the detectability of quadratic QNMs see \cite{yi2024nonlinear}. 


\section{Discussion}

In conclusion, SHG has been observed in numerical simulations of nonlinear BH ringdowns \cite{mitman2023nonlinearities, cheung2023nonlinear}. In quantum mechanics, second harmonics can produce non-classical states both in the parent and nonlinear modes, starting from initial coherent states \cite{mandel1982squeezing}. This can be understood in terms of particle creation and annihilation: nonlinearities affect particle number distributions, leading to states which classical theory cannot account for \cite{hillery1985conservation}. 

In principle, quantum features of such states could be directly measured using interferometric detectors. Then, one could set up a gravitational non-classicality witness and establish the quantum nature of gravity at the same level of rigor of quantum optics experiments. In practice, however, this is exceedingly hard due to the extremely weak coupling between GWs and matter. 


In quantum optics, it is well known that the output of nonlinear interactions is sensitive to the statistical features of the interacting fields. For instance, the efficiency of frequency doubling depends on the bunching or anti-bunching characteristics of the input field \cite{ekert1988second}. This suggests that strong gravitational nonlinearities could not only source non-classical GW states, but also function as efficient detectors of their quantum features. 

In this framework, we could envision an experiment for detecting the quantum nature of GWs occurring within the vicinity of a BH merger. As the BHs coalesce, nonlinear QNMs are excited, leading to the generation of non-classical states. The states from different modes then interact via their nonlinear dynamics, imprinting quantum statistical features on each other. Although some of this information will be dissipated due to the damped nature of QNMs, a fraction from low-damping modes can escape and be measured by interferometric detectors. It is conceivable that BH spectroscopy of nonlinear ringdowns could reveal signs of the quantum nature of gravitational radiation.
\\


\noindent \textbf{\textit{Note added:}} We have recently become aware that related ideas have been proposed in \cite{ manikandan2025squeezed}. 

\section*{Acknowledgments}
The author acknowledges  Antonio Zelaquett Khoury, George Svetlichny, Carlos Tomei, Kaled Dechoum, Luca Abrahão, Felipe Sobrero and Germain Tobar for conversations. We acknowledge support from the Coordenac\~ao de Aperfei\c{c}oamento de Pessoal de N\'ivel Superior - Brasil (CAPES) - Finance Code 001, Conselho Nacional de Desenvolvimento Cient\'ifico e Tecnol\'ogico (CNPq), Funda\c{c}\~ao de Amparo \`a Pesquisa do Estado do Rio de Janeiro (FAPERJ Scholarship No. E-26/200.251/2023 and E-26/210.249/2024), Funda\c{c}\~ao de Amparo \`a Pesquisa do Estado de São Paulo (FAPESP processo 2021/06736-5), the Serrapilheira Institute (grant
No. Serra – 2211-42299) and StoneLab.

\appendix

\section{Energy conservation and frequency doubling}\label{A_Energy_conservation}

Let $ a \ (a^{\dagger}) $, $ b \ (b^{\dagger}) $ denote annihilation (creation) operators for two Bosonic modes, satisfying the canonical commutation relations,
\begin{align}
    [a,a^{\dagger}] = [b^{\dagger}, b] = \mathds{1} 
\end{align}
\begin{align}
     [a, b] = [a^{\dagger}, b] = [a, b^{\dagger}] = [a^{\dagger}, b^{\dagger}] = 0
\end{align}
Define number operators $ N_{a} = a^{\dagger}a , N_{b} = b^{\dagger}b $. We want to find the most general Hermitian operator composed of at most third-order products of creation and annihilation operators that commutes with 
\begin{eqnarray}
    J = N_{a} + 2N_{b}
\end{eqnarray}
For convenience we list all the commutators with third-order operators in Table \ref{tab:commutators}.

\begin{table}[h]
\centering
\caption{\label{tab:commutators}Commutators with $J$.}
\begin{tabular}{@{}ccc}
\hline
$[a,J] = a$ & $[aa,J] = 2aa$ & $ [aaa,J] = 3aaa $ \\
$[a^{\dagger},J] = - a^{\dagger}$ & $ [bb,J] = 4bb $ & $ [aaa^{\dagger},J] = aaa^{\dagger} $ \\
$ [b,J] = 2 b $ & $ [ab,J] = 3ab $ & $ [aa^{\dagger}a,J] = aa^{\dagger}a $ \\
$ [b^{\dagger},J] = -2b^{\dagger} $ & $ [a^{\dagger}b,J] = a^{\dagger}b $ & $ [a^{\dagger}aa,J] = a^{\dagger}aa $ \\
 & $ [ab^{\dagger},J] = -ab^{\dagger} $ & $ [aa^{\dagger}a^{\dagger}, J] = -  aa^{\dagger}a^{\dagger} $ \\
 & $ [a^{\dagger}b^{\dagger},J] = -3a^{\dagger}b^{\dagger} $ & $ [a^{\dagger}aa^{\dagger},J] = - a^{\dagger}aa^{\dagger} $ \\
 & $ [a^{\dagger}a^{\dagger},J] = - 2 a^{\dagger}a^{\dagger} $ & $ [a^{\dagger}a^{\dagger}a,J] = - a^{\dagger}a^{\dagger}a $ \\ 
 & $ [b^{\dagger}b^{\dagger},J] = - 4 b^{\dagger}b^{\dagger} $ & $ [a^{\dagger}a^{\dagger}a^{\dagger},J] = - 3a^{\dagger}a^{\dagger}a^{\dagger} $ \\
\hline
\end{tabular}

\vspace{2mm}

\begin{tabular}{@{}ll}
$ [aab,J] = 4aab $ & $ [abb,J] = 5abb $ \\
$ [aab^{\dagger},J] = 0 $ & $ [abb^{\dagger},J] = abb^{\dagger} $ \\
$ [a^{\dagger}a^{\dagger}b,J] = 0 $ & $ [ab^{\dagger}b,J] = ab^{\dagger}b $ \\
$ [aa^{\dagger}b,J] = 2aa^{\dagger}b = 2a^{\dagger}ab+2b $ & $ [a^{\dagger}bb,J] = 3a^{\dagger}bb $ \\
$ [a^{\dagger}ab,J] = 2a^{\dagger}ab $ & $ [ab^{\dagger}b^{\dagger},J] = - 3ab^{\dagger}b^{\dagger} $ \\
$ [aa^{\dagger}b^{\dagger},J] = -2aa^{\dagger}b^{\dagger} = -2a^{\dagger}ab^{\dagger} - 2b^{\dagger} $ & $ [a^{\dagger}bb^{\dagger},J] = - a^{\dagger}bb^{\dagger} $ \\
$ [a^{\dagger}ab^{\dagger},J] = -2a^{\dagger}ab^{\dagger} $ & $ [a^{\dagger}b^{\dagger}b,J] = - a^{\dagger}b^{\dagger}b $ \\
$ [a^{\dagger}a^{\dagger}b^{\dagger},J] = -4 a^{\dagger}a^{\dagger}b^{\dagger} $ & $ [a^{\dagger}b^{\dagger}b^{\dagger},J] = -5 a^{\dagger}b^{\dagger}b^{\dagger} $ \\
\hline
\end{tabular}
\end{table}

\noindent Note that commutators with triple $ b$'s will have the same values as those with triple $ a$'s with an extra factor of 2.
The commutator defines a linear transformation $ T(\  \_ \ ) = [\ \_ \ ,J] $ from Hermitian to anti-Hermitian operators. By inspection, we can see the kernel of $ T $ is given by $ H_{ab} $ as defined in the main text.  
Energy conservation is sufficient to single out the frequency doubling Hamiltonian from the structure of eq.~\eqref{teukolsky}.


\section{Frequency doubling equations}\label{Freq_doubling_eqs}


Consider the Markovian open quantum system evolution for an operator $\mathcal{O} $, 
\begin{equation}
    \dot{\mathcal{O}} = i [H,\mathcal{O}] + \sum_{\nu} \left(  L_{\nu}^{\dagger} \mathcal{O} L_{\nu} - \frac{1}{2} L_{\nu}^{\dagger} L_{\nu}\mathcal{O} - \frac{1}{2} \mathcal{O} L_{\nu}^{\dagger} L_{\nu} \right) \ .
    \label{lindblad}
\end{equation}
We are interested in studying the evolution of the creation and annihilation operators of two Bosonic modes $ a, b $ evolving according to 
\begin{equation}
    H = H_{a} + H_{b} + H_{I} ,
\end{equation}
where 
\begin{equation}
    H_{a} = \omega a^{\dagger}a \ , H_{b} = 2\omega b^{\dagger} b \ , H_{I} = \kappa \left(  b^{\dagger}a^{2} + b a^{\dagger 2}  \right) 
\end{equation}
and 
\begin{equation}
    L_{a} = \sqrt{2\gamma_{a}} a \ , L_{b} = \sqrt{2\gamma_{b}} b \ .
\end{equation}

\noindent Denine the operators $ \alpha $ and $ \beta $ according to,
\begin{equation}
    \alpha = ae^{i\omega t} \ , \beta = b e^{2i\omega t}
\end{equation}
Note that $ \alpha $ and $ \beta $ have the same commutation relations as $ a, b$. 
Eq. \eqref{lindblad} for modes $ a $ and $ b $ then reads,
\begin{equation}
    \dot{\alpha} = -2i\kappa \beta \alpha^{\dagger} - \gamma_{a}\alpha 
\end{equation}

\begin{equation}
    \dot{\beta} = -i\kappa \alpha^{2} - \gamma_{b}\beta
\end{equation}

\section{Interaction between GW modes and cavity}\label{Optogravitational_Hamiltonian}

We briefly review the interaction Hamiltonian between a GW and a model GW detector and detail some calculations used in the main text. We follow the treatment outlined in \cite{pang2018quantum, guerreiro2020gravity, guerreiro2022quantum}.

\subsection{Hamiltonian}

Consider an optical cavity as a model for the GW detector. 
Starting from linearised Einstein's equations in Minkowski background, one can show that the interaction consists of a direct coupling between the optical modes and the GW perturbations \cite{pang2018quantum}.
The total Hamiltonian for the GW + optical modes is given by
\begin{equation}
    H = H_{F} + H_{O}
\end{equation}
where $ H_{F} $ are the free field Hamiltonians (to be discussed later) and $ H_{O} $ describes the interaction between the GW modes and the detector. 
For a $ + $ polarized GW propagating perpendicular to the cavity axis the interaction is given by 
\begin{equation}
H_{O} = - \dfrac{\omega_{c}}{4} c^{\dagger} c \int \dfrac{d \bm{k}}{\sqrt{(2\pi)^{3}}} \left(   \sqrt{\dfrac{8\pi G}{k}} \mathfrak{b}_{\bm{k}} + h.c.  \right) \ ,
\label{H_int_canonical}
\end{equation}  
where $ k = \vert \bm{k} \vert = \omega_{k} $ is the GW frequency for the mode $ \bm{k} $, the operator $  \mathfrak{b}_{k}^{\lambda} $ ($ \mathfrak{b}_{\bm{k}}^{\lambda \dagger} $) is the canonical graviton annihilation (creation) operator and $ \omega_{c} $ is the cavity frequency with photon annihilation (creation) operator $ c \ (c^{\dagger}) $.
To obtain a discrete version of \eqref{H_int_canonical} we introduce a quantization volume $ V $ and note that $ \left[ \sqrt{8\pi G / k} \right] = L^{3/2} $, where $ L$ denotes dimension of length. Moreover, $ \left[ d \bm{k} \right] = L^{-3} $. The graviton annihilation and creation operators then have dimension $ \left[ \mathfrak{b}_{\bm{k}} \right] = L^{3/2} $. Define $ \mathfrak{b}_{\bm{k}} = \sqrt{V}  \bm{b}_{\bm{k}} $, where $  \bm{b}_{\bm{k}} $ is dimensionless.  The discrete limit $ d \bm{k} \rightarrow 1 / V $, $ (2\pi)^{-3/2}\int \rightarrow \sum $ leads to, 
\begin{equation}\label{eq:H_I}
H_{O} = - \dfrac{\omega_{c}}{4} c^{\dagger} c \sum_{\bm{k}} \left(   \sqrt{\dfrac{8\pi G}{kV}} \bm{b}_{\bm{k}} + h.c. \right).
\end{equation}
Defining the single graviton strain for mode $ \bm{k} $ as $ f_{\bm{k}} =  \sqrt{8\pi G/(kV)} $, then the coupling constant of a GW mode $ \bm{k} $ with the cavity electromagnetic field is $ g_{\bm{k}} = \omega_{c}  f_{\bm{k}} / 4 $, which has dimension of frequency.
It is also convenient to introduce the dimensionless coupling $ q_{\bm{k}} =  g_{\bm{k}} / \omega_{k} $.

The discretized free-field Hamiltonian reads
\begin{equation}\label{eq:H_0}
H_{F} = \omega_{c} c^{\dagger} c + \sum_{\bm{k}} \omega_{k} \bm{b}_{\bm{k}}^{\dagger} \bm{b}_{\bm{k}} \,.
\end{equation}

In the interaction picture, the time evolution is given by 
\begin{equation}
U(t) = e^{-i H t} = \prod_{\bm{k}} U_{\bm{k}}(t)\,,
\label{total_U}
\end{equation}
where,
\begin{equation}
U_{\bm{k}} (t)  = e^{iB_{k}(t) (c^{\dagger}c)^{2}} e^{q_{\bm{k}} c^{\dagger} c  (\gamma_{k} \bm{b}^{\dagger}_{k} -\gamma_{k}^{*} \bm{b}_{k})} \,,
\label{Uk}
\end{equation}
with 
\begin{equation}
    \gamma_{k} = (1 - e^{-i\omega_{k} t}) \label{gammas} 
\end{equation}

\begin{equation}
    B_{k}(t) = q_{k}^{2} \left(   \omega_{k} t - \sin \omega_{k} t  \right) \label{Bks}
\end{equation}
and GW states evolve according to \cite{brandao2020entanglement},
\begin{equation}
\vert \Psi(t) \rangle = \prod_{\bm{k}} e^{-i \bm{b}_{\bm{k}}^{\dagger} \bm{b}_{\bm{k}} \omega_{\bm{k}} t} \vert \Psi \rangle \,.
\end{equation}

\subsection{Multimode vacuum corrections and few-mode Hamiltonian}

The evolution operator \eqref{total_U} contains contributions from all modes of the gravitational field, most of which are in the vacuum state. Following \cite{guerreiro2020gravity}, we now show that the contribution due to empty modes can be safely neglected, as expected.
The evolution of the cavity annihilation operator is given by,
\begin{eqnarray}
c(t) &=& \left( \prod_{k} U_{k}(t)   \right)^{\dagger} c \left( \prod_{k} U_{k}(t)   \right) \nonumber \\
&=& \prod_{k} e^{iB_{k}(t) c^{\dagger}c} e^{i B_{k}(t)/2} e^{q_{k}c^{\dagger}c\left( \gamma_{k}\bm{b}^{\dagger} - \gamma_{k}^{*} \bm{b}_{k}   \right)} \ c  \nonumber \\
&=& \prod_{k} e^{iB_{k}(t) c^{\dagger}c} e^{i B_{k}(t)/2} D(q_{k}\gamma_{k}) \ c
\label{annihilation_op}
\end{eqnarray}
where $ D(q_{k}\gamma_{k}) $ is the GW displacement operator. Consider now the GW field to be in the vacuum state
\begin{eqnarray}
\vert 0 \rangle = \prod_{k} \vert 0_{k} \rangle 
\end{eqnarray}
where $ \vert 0_{k} \rangle  $ denotes the vacuum state in mode $ k $. The annihilation operator of an optical mode interacting with such gravitational vacuum can then be written as 
\begin{eqnarray}
c(t) = e^{i \mathcal{F}(t) c^{\dagger} c} e^{i\mathcal{F}(t)/2} \mathcal{G}(t) \ c
\end{eqnarray}
where 
\begin{eqnarray}
\mathcal{F}(t) = \sum_{k} B_{k}(t)
\end{eqnarray}
and 
\begin{eqnarray}
 \mathcal{G}(t) = \prod_{k} \langle 0_{k} \vert D(q_{k}\gamma_{k}) \vert 0_{k} \rangle = \exp \left[ -\dfrac{1}{2} \sum_{k} q_{k}^{2}\vert \gamma_{k}\vert^{2} \right]
\end{eqnarray}
On physical grounds, these expressions must be cut-off at a maximum and minimum frequencies $ \omega_{k} = \vert k \vert $ defining the interval over which the detector is sensitive \cite{parikh2021b}; consider for the purpose of illustration these infrared and ultraviolet cut-offs as the Hubble and Planck energies, $ E_{\rm IR} $ and $ E_{\rm pl}$, respectively. Recovering the continuous limit gives us, up to numerical factors of order one,
\begin{eqnarray}
 \mathcal{F}(t) 
 = \int \frac{d^{3} \bm{k}}{\sqrt{(2\pi)^{3}}} \,
   2\omega_{c}^{2} \left( \frac{8\pi G}{k^{3}} \right)
   \left( \omega_{k} t - \sin \omega_{k} t \right) \nonumber \\
 \quad \approx  
   \left( \frac{\omega_{c}}{E_{\rm pl}} \right)^{2}
   \left( \int_{E_{\rm IR}}^{E_{\rm pl}} dk \right) t 
   \approx \left( \frac{\omega_{c}}{E_{\rm pl}} \right) \omega t ,
\label{phase_noise}
\end{eqnarray}
where we have used $ E_{\rm pl} \gg E_{\rm IR} $, considered large times $ t \gg E_{\rm IR}^{-1} $ and neglected the bounded term $ \sin \omega_{k} t  $.
Corrections to the optical annihilation operator due to the phase $ \mathcal{F}(t) $ are then on the order of $ (\omega_{c} / E_{\rm pl}) $ and only become relevant for optical frequencies close to the Planck energy.
Similarly,
\begin{eqnarray}
\mathcal{G}(t) 
= \exp\!\left[ -\frac{1}{2} \int \frac{d^{3} \bm{k}}{\sqrt{(2\pi)^{3}}} 
   \, \omega_{c}^{2} \left( \frac{8\pi G}{k^{3}} \right) 
   \vert \gamma_{k} \vert^{2} \right] \nonumber \\[2mm]
\approx \exp\!\left[ - 2 \left( \frac{\omega_{c}}{E_{\rm pl}} \right)^{2} 
   \int_{E_{\rm IR}}^{E_{\rm pl}} \frac{dk}{k} \right] \nonumber \\[2mm]
= \exp\!\left[ - 2 \left( \frac{\omega_{c}}{E_{\rm pl}} \right)^{2} 
   \ln\!\left( \frac{E_{\rm pl}}{E_{\rm IR}} \right) \right] .
\end{eqnarray}

\noindent where we have considered the worst-case approximation $ \vert \gamma_{k} \vert^{2} = 2 \left( 1 - \cos \Omega_{k} t \right) \sim 4 $. Notice that the ratio of ultraviolet to infrared cut-offs is $ E_{\rm pl} / E_{\rm IR} \approx 10^{62} $, giving a correction to $ c(t) $ approximately proportional to $ (\omega_{c} / E_{\rm pl})^{2} $.

Instead of the vacuum, we could consider all modes to be populated by a thermal state at about \SI{1}{K}, the expected temperature for the cosmic GW background \cite{allen1996stochastic}. This would not alter $ \mathcal{F}(t) $, which is state-independent, so the estimates in \eqref{phase_noise} remain. The $ \mathcal{G}(t) $ term, however, would acquire a correction factor at most $ e^{-q_{\rm pl}^{2} (2 \bar{n} + 1)} $, with $ \bar{n} = 1 / (e^{\hbar \Omega_{k} / k_{B}T} - 1) $ and $ q_{\rm pl}
$ the GW coupling strength at the Planck frequency \cite{guerreiro2020gravity}. For a GW mode of \SI{10}{Hz}, peak of the expected cosmic GW background spectrum, this correction factor amounts to $ \approx e^{-10^{-47}} $. 

All in all, these estimates show that the effect of modes which are not populated by states with a large mean number of gravitons upon optical observables is negligible. In other words, decoherence due to the gravitational vacuum, or even due to the thermal background of GWs is very weak, which is consistent with previous results \cite{dyson2014graviton,  blencowe2013effective}. 

\subsection{Few mode Hamiltonian}

From now on, we will take a \textit{few mode} approximation and only consider the terms in the interaction Hamiltonian for which the GW mode is non-empty. For our parent and nonlinear ringdown modes we will consider 
\begin{equation}
    H_{O} = \omega_{c} c^{\dagger}c + \omega_{a}a^{\dagger}a + \omega_{b}b^{\dagger}b - c^{\dagger}c \left(  g_{a}Q_{a} + g_{b} Q_{b}   \right)
\end{equation}
where the GW parent and nonlinear modes are denoted as $ a $ and $ b $, with frequencies $ \omega_{a} $ and $ \omega_{b} = 2\omega_{a} $, respectively. We have defined amplitude quadrature operators for modes $ a $ and $ b $,
\begin{equation}
    Q_{a} = a + a^{\dagger} \ , \ Q_{b} = b + b^{\dagger} 
\end{equation}
Recall also the definition of phase operators as,
\begin{equation}
    P_{a} = i(a^{\dagger} - a) \ , \ P_{b} = i(b^{\dagger} - b)
\end{equation}
in accordance do Eq. \eqref{quadratures_times}.
Note the two modes couple differently to the GW detector; we have,
\begin{equation}
    \frac{q_{a}}{q_{b}} = \left(\frac{\omega_{b}}{\omega_{a}}\right)^{3/2} = 2\sqrt{2}
\end{equation}
The total unitary evolution in the interaction picture reads,
\begin{equation}
    U(t) = e^{-iB(t)(c^{\dagger}c)^{2}} e^{q_{a}c^{\dagger}c(\gamma_{a}a^{\dagger} - \gamma_{a}^{*}a)} e^{q_{b}c^{\dagger}c(\gamma_{b}b^{\dagger} - \gamma_{b}^{*}b)}
\end{equation}
with $ B(t) = \sum_{k = a,b} B_{k}(t) $, and $ \gamma_{k}, B_{k}(t) $ are defined in eqs. \eqref{gammas} and \eqref{Bs} above. 

\subsection{Optical density matrix}

Consider an initial optical + GW state of the form 
\begin{equation}
    \vert \psi_{0} \rangle = \vert \Phi \rangle \vert \Psi \rangle
\end{equation}
where $ \vert \Psi \rangle $ is the GW state and $ \vert \Phi \rangle $ is a general optical pure state given by
\begin{equation}
    \vert \Phi \rangle = \sum_{n} a_{n} \vert n \rangle
\end{equation}
We are interested in calculating how the elements of the reduced optical optical density matrix depend on the initial GW state. From measurements of the optical density matrix, we can then obtain information on the parent and nonlinear GW modes, and in particular some of their correlation functions. 

The general optical density matrix element $[\rho_{c}(t)]_{nm} $ in the number basis reads
\begin{equation}
    [\rho_{c}(t)]_{nm} = a_{n}a^{*}_{m} e^{-i \delta B(t)} e^{-i\omega_{c}\Delta t} \ \mathcal{C}_{\Delta}(t)
\end{equation}
where $ \delta = (n^{2} - m^{2}) \ , \Delta = (n-m) $ and we have defined the \textit{correlation function} $ \mathcal{C}_{\Delta}(t)$, 
\begin{eqnarray}
    \mathcal{C}_{\Delta}(t) &=& \langle \Psi(t) \vert e^{\Delta q_{a}(\gamma_{a}a^{\dagger} - \gamma_{a}^{*}a))} e^{\Delta q_{b}(\gamma_{b}b^{\dagger} - \gamma_{b}^{*}b))} \vert \Psi(t) \rangle = \\
    & = & \langle \Psi \vert e^{\Delta q_{a}(\gamma_{a}^{*}a^{\dagger} - \gamma_{a}a))} e^{\Delta q_{b}(\gamma_{b}^{*}b^{\dagger} - \gamma_{b}b))} \vert \Psi \rangle
    \label{correlator}
\end{eqnarray}

Notice that the dependence of $ [\rho_{c}(t)]_{nm} $ on the initial GW state $ \vert \Psi \rangle $ is entirely contained in $ \mathcal{C}_{\Delta}(t)$, with the pre-factor being a known time-dependent function which is the same for any GW state. Therefore, if we know the initial optical state $ \vert \Phi \rangle $ (and its decomposition in terms of the number basis) and the GW modes frequencies, by measuring the optical density matrix we can obtain information on the parent and nonlinear GW correlations. From now on we will treat $ [\rho_{c}(t)]_{nm} $ and $ \mathcal{C}_{\Delta}(t)$ interchangeably, in the sense that knowing one implies knowledge of the other. Measuring the optical density matrix therefore allows for measurements of the GWs' correlations. Note that we can generalize \eqref{correlator} to mixed states. For a general initial mixed GW state $ \sigma  $ (unentangled with the optical mode) we have,
\begin{equation}
    \mathcal{C}_{\Delta}(t) = \mathrm{Tr} \left( \sigma   e^{\Delta q_{a}(\gamma_{a}^{*}a^{\dagger} - \gamma_{a}a))} e^{\Delta q_{b}(\gamma_{b}^{*}b^{\dagger} - \gamma_{b}b))}  \right)
\end{equation}

Expanding $ \mathcal{C}(t) $ to quadratic order in the couplings $ q_{a}, q_{b} $ we get,
\begin{eqnarray}
\mathcal{C}_{\Delta}(t) \approx 1 
- i q_{a} \Delta \left( \sin \omega_{a}t \langle X_{a} \rangle 
+ (1 - \cos \omega_{a}t) \langle Y_{a} \rangle \right) \nonumber \\[1mm]
- \frac{q_{a}^{2}\Delta^{2}}{2} \Big( 
\sin^{2}\omega_{a}t \, \langle X_{a}^{2} \rangle 
+ (1 - \cos \omega_{a}t)^{2} \langle Y_{a}^{2} \rangle  \nonumber \\[1mm]
\qquad\qquad\quad + \sin \omega_{a}t (1 - \cos \omega_{a}t) 
\langle X_{a}Y_{a} + Y_{a}X_{a} \rangle \Big) \nonumber \\[1mm]
- i q_{b} \Delta \left( \sin \omega_{b}t \langle X_{b} \rangle 
+ (1 - \cos \omega_{b}t) \langle Y_{b} \rangle \right) \nonumber \\[1mm]
- \frac{q_{b}^{2}\Delta^{2}}{2} \Big( 
\sin^{2}\omega_{b}t \, \langle X_{b}^{2} \rangle 
+ (1 - \cos \omega_{b}t)^{2} \langle Y_{b}^{2} \rangle  \nonumber \\[1mm]
\qquad\qquad\quad + \sin \omega_{b}t (1 - \cos \omega_{b}t) 
\langle X_{b}Y_{b} + Y_{b}X_{b} \rangle \Big) \nonumber \\[1mm]
- q_{a}q_{b}\Delta^{2} \Big( 
\sin \omega_{a}t \sin \omega_{b}t \langle X_{a} X_{b} \rangle 
+ \sin \omega_{a}t (1 - \cos \omega_{b}t) \langle X_{a} Y_{b} \rangle \nonumber \\[1mm]
\qquad\qquad\quad + \sin \omega_{b}t (1 - \cos \omega_{a}t) \langle Y_{a} X_{b} \rangle 
+ (1 - \cos \omega_{a}t)(1 - \cos \omega_{b}t) \langle Y_{a} Y_{b} \rangle \Big) . \nonumber \\
\end{eqnarray}

\noindent We see this contains terms corresponding to each one of the fourteen second-order amplitude and phase correlators for modes $a $ and $ b $ multiplied by independent functions of time. By collecting the value to $ \mathcal{C}_{\Delta}(t) $ (or equivalently $ [\rho_{c}(t)]_{nm} $) at distinct times, taking care to avoid aliasing of the coefficient functions, we can obtain the correlators necessary to verify the Duan criteria for the parent and nonlinear GW modes.

\section{Parent mode quadrature in frequency doubling}\label{SHG_solutions}

We are interested in the behavior of $ \beta(t) $ for small interaction times. To linear order in $ t $ \cite{kozierowski1977quantum, mandel1982squeezing},
\begin{eqnarray}
    \beta(t) &=& \beta(0) + t \dot{\beta}(0) + \frac{t^{2}}{2}\ddot{\beta}(0) + ... \\
    &=& \beta(0) + t \left( - i\kappa \alpha^{2}(0) - \gamma_{b}\beta(0) \right) + \mathcal{O}(t^{2}) 
\end{eqnarray}
Consider the initial state for modes $ a, b $ is $ \vert \Psi\rangle_{a} \vert 0 \rangle_{b} $. Then, $ \langle \beta(0) \rangle = \langle \beta^{\dagger}(0) \rangle = 0  $, and we can write,
\begin{equation}
    \langle \beta(t) \rangle \approx -i\kappa t \langle \alpha^{2}(0) \rangle
\end{equation}
Define the field quadratures
\begin{equation}
    Q_{z}(t) = z e^{i\omega_{z}t} + z^{\dagger} e^{-i\omega_{z}t} \ , \ 
    P_{z}(t) = i\left(  z^{\dagger} e^{-i\omega_{z}t} - z e^{i\omega_{z}t} \right)
    \label{quadratures_times}
\end{equation}
where $ z = a,b $. To first order in $ t $ we have,
\begin{equation}
    \langle Q_{a}(t) \rangle \approx \frac{\kappa t}{2} \Bigl< Q_{a}(0)P_{a}(0) + P_{a}(0)Q_{a}(0) \Bigr> 
\end{equation}

A Gaussian state is completely determined by the mean values of the field quadratures and their covariance matrix $ \mathds{V} $. For a single mode state with quadrature operators $ X,Y $ we have $ \mathds{V}_{XY} = \frac{1}{2}  \Bigl< XY + YX\Bigr> - \langle X \rangle \langle Y \rangle $. The most general Gaussian state corresponds to a displaced, rotated, squeezed state, and has covariance matrix \cite{weedbrook2012gaussian},
\begin{equation}
    \mathds{V} = (2\bar{n}+ 1) \mathbf{R}(\phi) \mathbf{S}(2r) \mathbf{R}(\phi)^{T}
    \label{covariance}
\end{equation}
where $ \mathbf{R}(\phi) $ is a $ 2 \times 2 $ rotation matrix, $ \mathbf{S}(2r) = \mathrm{diag}(e^{2r}, e^{-2r}) $ is the squeezing matrix with parameter $ r $, and $ \bar{n}$ is the thermal state mean occupation number. From the definition of the covariance matrix and Eq. \eqref{covariance} we have,
\begin{equation}
    \frac{1}{2}  \Bigl< XY + YX\Bigr> = (2\bar{n}+ 1)\sinh r \sin \phi \cos \phi + \langle X \rangle \langle Y \rangle 
\end{equation}
where $ \langle X \rangle = \vert h \vert \cos \theta $ and $ \langle Y \rangle = \vert h \vert \sin \theta  $ are the mean quadratures of a coherent state with amplitude $ he^{i\theta} $.
When $ \bar{n} = 0 $ we get the most general pure Gaussian state, corresponding to a rotated squeezed-coherent state,
\begin{equation}
    \vert \Psi \rangle = D(h)R(\phi)S(r)\vert 0 \rangle
\end{equation}
where $ D, R, S $ denote the displacement, phase rotation and squeezing operators, respectively.

\bibliography{main}

\end{document}